\documentclass{article}

\usepackage{amsmath,amsfonts,amssymb,bm}
\usepackage{mathptmx}
\usepackage{newtxtext}
\usepackage{newtxmath}
\usepackage{authblk}
\usepackage{natbib}
\usepackage[french,english]{babel}
\usepackage[utf8]{inputenc}
\usepackage{graphicx}
\usepackage[T1]{fontenc}
\usepackage{hyperref}

\title{ARPEGE Cloud Cover Forecast Post-Processing with Convolutional Neural Network}

\begin{document}

    \author[1]{Florian Dupuy \thanks{Corresponding author: \texttt{florian.dupuy@irt-saintexupery.com}}}
    \author[2]{Olivier Mestre}
    \author[3]{Mathieu Serrurier}
    \author[1]{Mohamed Chafik Bakkay}
    \author[1]{Valentin Kivachuk Burd\'{a}}
    \author[4]{Naty Citlali Cabrera-Guti\'errez}
    \author[4]{Jean-Christophe Jouhaud}
    \author[1]{Maud-Alix Mader}
    \author[1]{Guillaume Oller}
    \author[2]{Michaël Zamo}

    \affil[1]{Institut de Recherche Technologique Saint-Exupéry, Toulouse, France}
    \affil[2]{Météo-France, Direction des Opérations pour la Production, 42 avenue Gaspard Coriolis, 31057 Toulouse cedex 07, France and CNRM/GAME, Météo-France/CNRS URA 1357, Toulouse, France}
    \affil[3]{IRIT, Université Paul Sabatier, Toulouse, France}
    \affil[4]{CERFACS, Toulouse, France}

\maketitle

\begin{abstract}
Cloud cover is crucial information for many applications such as planning land observation missions from space. It remains however a challenging variable to forecast, and Numerical Weather Prediction (NWP) models suffer from significant biases, hence justifying the use of statistical post-processing techniques. In this study, ARPEGE (Météo-France global NWP) cloud cover is post-processed using a convolutional neural network (CNN). CNN is the most popular machine learning tool to deal with images. In our case, CNN allows the integration of spatial information contained in NWP outputs.
We use a gridded cloud cover product derived from satellite observations over Europe as ground truth, and predictors are spatial fields of various variables produced by ARPEGE at the corresponding lead time. We show that a simple U-Net architecture produces significant improvements over Europe. Moreover, the U-Net outclasses more traditional machine learning methods used operationally such as a random forest and a logistic quantile regression. 
We introduced a weighting predictor layer prior to the traditional
U-Net architecture which produces a ranking of predictors by
importance, facilitating the interpretation of the results. Using $N$
predictors, only $N$ additional weights are trained which does not
impact the computational time, representing a huge advantage compared
to traditional methods of ranking (permutation importance, sequential
selection, \ldots).
\end{abstract}

%
%
%

%

\section{Introduction}

The highly chaotic nature of the atmospheric dynamic makes numerical weather prediction (NWP) a difficult task and errors are frequent. Forecast errors are caused by a combination of inaccurate forcing (initial/boundary conditions) and incomplete mathematical representation of phenomena. Cloud representation in NWP models is a crucial issue, due to many interactions with dynamics, radiation, surface energy budget and aerosols. However, cloudiness remains one of the most difficult parameters to predict \citep{haiden2015}. \citet{haiden2016} demonstrated that the skill of 24-h ECMWF total cloud cover (TCC) forecasts verified against a set of European stations improved little over the last decade. In comparison with other variables, such as 6-h accumulated precipitation, geopotential, 2-m temperature or 10-m wind speed, the skill of NWP TCC forecasts is low \citep{kohler2005}. \citet{morcrette2012} categorized cloud errors to be one of three basic types: frequency of occurrence, amount when present and timing errors in terms of the diurnal cycle/time-of-day. Important known biases are related to the representation of low-level clouds and fog \citep{kann2010, roman2016, steeneveld2015}, mostly related to timing errors in the formation/disappearance and spatial extend, and convection cumulus clouds.

Most national weather services add a post-processing step, also known as model output statistics (MOS), in order to improve their forecasts. Numerous methods were successfully used: logistic regression \citep{walker1967, hamill2004}, random forest \citep{breiman2001, zamo2016}, \ldots. However, there is no clue to knowing which method will yield the best results on a given problem. The only way to pick the best method is to compare results on the given problem.

Although plenty of studies exist on weather forecasts postprocessing, few concern cloud cover. \citet{hemri2016} have postprocessed ensemble cloud cover forecasts from the European Center for Medium-range Weather Forecasts (ECMWF). The discrete cloud cover was calculated (classification problem) at several stations’ locations across the globe using either a multinomial logistic regression or a proportional odds logistic regression model. \citet{baran2020} extended that study by comparing other methods including random forests and neural networks (NNs). The NN showed the best performances.

NNs are increasingly used in a wide scope of applications related to atmospheric science (see \citet{gardner1998, dueben2018, boukabara2019} for an overview). Convolutional neural networks \citep{lecun2015} (CNNs) are a specialized kind of neural networks processing grid-like data including images. The goal of a CNN is to extract hierarchical features from the input image through convolutions. That makes it a suitable tool for working with geophysical data in order to extract spatial features.
	
The atmospheric research community has already taken advantage of CNN's ability (see \citet{reichstein2019} for an overview). Most of the applications deal with images, for example from satellite observations to create cloud masks or derive rainfalls \citep{dronner2018, moraux2019}, or from pictures for weather classification \citep{elhoseiny2015}.

Generally, CNNs using NWP data as predictors are used to produce either a classification or a pointwise regression, meaning that the CNN produces a zero dimension result from two dimensional data. For example to correct the precipitation forecast integrated over a region, to estimate if a thunderstorm will produce large hailstones, or to predict if a storm will generate a tornado \citep{pan2019, gagne2019, lagerquist2019a}. Few NWP postprocessing using CNNs has been performed on a grid scale. \citet{vandal2018} performed a statistical downscaling of climatic precipitation simulations, using high resolution topography information. \citet{bano2019} had a similar approach with an additional comparison with standard methods demonstrating the superiority of CNNs. \citet{lagerquist2019} used CNNs to automatically generate front maps from the North American Regional Reanalysis. Again, CNNs outperformed standard methods.
	
In this study, we evaluate the ability of CNNs to post-process ARPEGE cloud cover forecasts on a grid scale. The area and the dataset are presented in the next section. The section \ref{sec:sec3} is dedicated to the presentation of the machine learning algorithms used and of the forecast evaluation methodology. Results are presented and discussed in the section \ref{sec:sec4}, including a discussion on the importance of predictors based on a novel CNN-based method.


\section{Data\label{sec:sec2}}
    Our dataset is composed of an analysis of total cloud cover (cf. section 2.\ref{sec:sec2-2}) and modeled data: ARPEGE NWP forecasts concerning weather fields (cf. section 2.\ref{sec:sec2-3}) and SURFEX for terrain data (cf. section 2.\ref{sec:sec2-4}). The analysis is considered as the ground truth and is used to evaluate the ARPEGE and post-processed TCC forecasts, as well as to train the machine learning algorithms. All the data were produced on a regular grid of $0.1^{\text{o}} \times 0.1^{\text{o}}$ over an enlarged region of Europe (cf. section 2.\ref{sec:sec2-1}). Only data at 15h00 UTC are considered across a two years period (2017 -- 2018). After removing the dates for which data are missing, there are 662 days left.

    
    \subsection{Area description\label{sec:sec2-1}}
    
\begin{figure}[h]
 \centerline{\includegraphics[width=\linewidth]{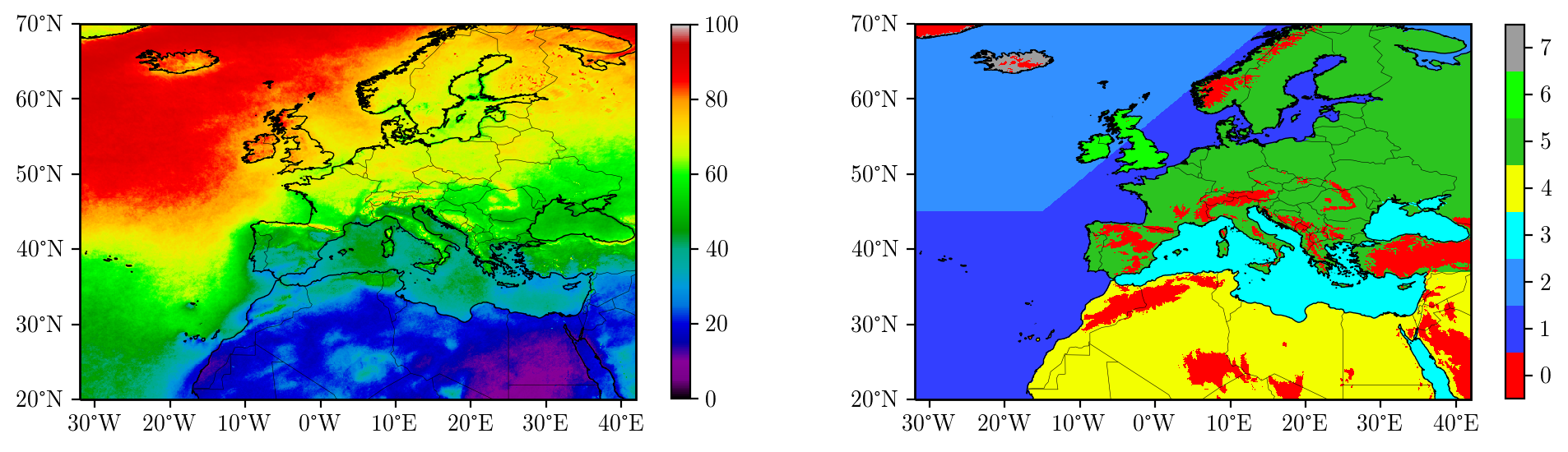}}
  \caption{Mean TCC (\%) from the analysis over the 2017--2018 period (left) and regions used as stratification for the evaluation of forecasts (right): 0 for the mountains (altitude over 800 m), 1 for the southern and coastal part of the Atlantic Ocean, 2 for the northern part of Atlantic Ocean, 3 for the Mediterranean, Black and Red seas, 4 for Africa and Middle East, 5 for continental Europe, 6 for British Isles and 7 for Iceland and Greenland.}\label{fig1}
\end{figure}

    We focus our study on a region extending from 20$^{\text{o}}$N to 70$^{\text{o}}$N in latitude and from 32$^{\text{o}}$W to 42$^{\text{o}}$E in longitude (see figure \ref{fig1}). This includes many different climates, from the very dry and sunny Sahara desert to the very cloudy polar conditions of Iceland. 

    These heterogeneous conditions inevitably lead to different cloud cover characteristics. For example, oceans, which represent a large part of the domain, are characterized by overall higher cloud fractions \citep{king2013}. Big mountains, known to have thicker cloud covers and with a higher occurrence \citep{barry2008}, such as the Alps, the Atlas, the Pyrenees, the Carpathian mountains or Turkish mountains are also present in the area.


    \subsection{Analysis of TCC\label{sec:sec2-2}}
    
    The analysis of TCC, produced by Météo-France, is derived from geostationary satellite observations. The TCC is calculated based on cloud type classification. The value of a given grid cell corresponds to the mean value on an approximate 30-km radius circle area to approach observations values reported by a human observer. 

    Because cirrus are semi-transparent, the TCC associated with an overcast sky of cirrus (with no other types of cloud) is fixed to a value of 50 \%. This results in a tri-modal distribution, with local maxima at 0, 50 and 100 \%.


    \subsection{ARPEGE data\label{sec:sec2-3}}
    
    ARPEGE (Action de Recherche Petite Echelle Grande Echelle) is the global operational NWP system operated by Météo-France \citep{courtier1991}. ARPEGE forecasts run with a time step of 9 minutes on a stretched grid allowing a 7.5 km resolution grid mesh over France. The vertical discretization is performed on 105 levels, with the lowest one at 10 m. ARPEGE is initialized by a 4D-Var data assimilation scheme.

    We used the operational weather forecasts of the years 2017 and 2018. The version of the model did not evolve very much during that period, making the forecasts consistent during the 2-year period. The data are produced on a regular $0.1^{\text{o}} \times 0.1^{\text{o}}$ latitude/longitude grid on a domain encompassing Europe, North Africa and part of the Atlantic Ocean (see figure \ref{fig1} for the extent of the region). Only the $+15$ h forecasts from the daily simulations run at 00h00 UTC are used (forecasts valid at 15h00 UTC). At this time, corresponding to early afternoon over Europe and Africa (the region is crossed by 5 time zones due to its large longitudinal extension), convection starts to increase making the cloud forecast difficult.

    ARPEGE calculates different cloud-related variables (see \citet{seity2013} for a detailed description of cloud representation in ARPEGE). Firstly, cloud fractions (CF) are calculated for each cell (3D variable). They are then interpolated on altimetric coordinates at several levels above ground level. Secondly, vertical integrated clouds are calculated over 3 layers of the troposphere: low-level, mid-level and high-level CC (2D variables). Thirdly, convective cloud cover is calculated. And finally, the TCC over the whole column is calculated from the previous cloud variables. This is the ARPEGE forecast to be compared to the TCC analysis. 

    These cloud variables are used as predictors to calculate the TCC with the machine learning algorithms. Other variables from the ARPEGE forecasts are used as predictors: fundamental meteorological variables such as temperature and relative humidity (at several levels), sea level pressure, precipitations and winds; fluxes (radiative and thermal); atmosphere stability-related variables, such as boundary layer height, convective available potential energy (CAPE) or vertical difference of temperature.


    \subsection{Terrain data\label{sec:sec2-4}}
	
	In order to incorporate spatial context, topography and information on the type of the soil (proportions of nature, town, sea and land water bodies for each grid cell) is added to the list of predictors. We use the topography from the ARPEGE simulations and the types of soil come from the SURFEX model \citep{lemoigne2009}. The table \ref{tab1} summarizes the list of predictors used.


\section{Methods\label{sec:sec3}}
    
    
    In order to establish the score baseline, two methods already used in operations have been tested on our dataset: linear quantile regression (LQR), already used to compute ARPEGE MOS of total cloud cover over the globe, and block-MOS random forests (RF) described in \citet{zamo2016} for wind speed post-processing.
        
    \subsection{LQR\label{sec:sec3-1}}
    
    In this approach, the median of the target variable is modelled as a linear function of a set of covariates. Quantile regression (QR) makes no assumption on the shape of the distribution of the outcome \citep{koenker1978}. We discovered during preliminary studies (not shown here) that for linear methods, modelling the conditional median was more efficient than conditional mean: due to the peculiar bimodal distribution of observed and predicted TCC, Ordinary Least Squares regression would always fail to predict 0 and 100 \% values. This is simply due to forecast errors. The conditional mean of observed TCC values is never equal to 0 (resp. 100 \%) given raw predicted TCC values of 0 (resp. 100 \%), while conditional median of observed TCC is less prone to this phenomenon. Here, regressions are estimated separately for every grid point and lead time.

    For numerical estimation of the linear coefficients, we take advantage of the lqm function of the R lqmm package \citep{geraci2014b}. The function maximizes the log-likelihood of a Laplace regression. This is equivalent to the minimization of the weighted sum of absolute residuals \citep{koenker1978}. We faced many numerical problems when estimating our quantile regression: the current operational ARPEGE MOS application required estimating $20 \times 10^6$ equations corresponding to the number of grid points times the number of lead times -- thus requiring a very robust estimation procedure. We found out that the optimization algorithm based on the gradient of the Laplace log-likelihood implemented in lqm function \citep{bottai2015} met our requirements in terms of robustness. Lastly, variable selection is performed using the BIC \citep{schwarz1978} criterion and a simple backward procedure.

    \subsection{Random forest\label{sec:sec3-2}}
    
    Random forest \citep{breiman2001} is a classical machine learning technique. In a regression context, RF consists of averaging the output of several regression trees \citep{breiman1984} whose principle is recalled hereinafter. For a single regression tree, the regression function is built by iteratively splitting the target variable into two subsets. Splitting is done by looking for some optimal threshold over the set of quantitative explanatory variables. The splitting variable and the corresponding threshold is chosen so that the two subsets of response values have minimum intra-group variance (and maximum inter-group variance). Classically, a split is called a node, and final subsets are called leaves. Predicted values are simply the average of the response data within leaves. Depth of regression trees may be controlled via a parameter such as maximum number of nodes, or minimum number of observations in leaves. In RF, each tree is built according to two randomization schemes: firstly, training samples are bootstrapped. Secondly, during the construction of trees, at each node, a set of potential splitting variables is randomly selected among the set of explanatory variables. This randomization aims at building more independant trees. Each individual tree built this way would perform less well than a traditional regression tree. But the averaging response of those rather sub-optimal but much more independant trees reduces variance of errors without increasing bias \citep{breiman2001}.

    In our application, since we compute MOS across a large grid (more than 250000 gridpoints), we adopt a block-MOS procedure as in \citet{zamo2016}, that is to build a single random forest for groups of $3 \times 3$ gridpoints, pooling data of the corresponding gridpoints. Latitude and longitude are added as additional predictors, since some gridpoints may exhibit different behaviour within the block. Compared to pointwise training, this procedure has two advantages: firstly, it limits the number of forests to build during training, thus limiting the corresponding data to load and store into memory during operations, and secondly, it enhances the performances, since training is computed on far more data, as shown in \citet{zamo2016}. Preliminary study (not shown here) demonstrate that 200 trees are enough to ensure good performances, and we test whether shallower trees (maximum number of nodes $= 350$, model RF$_{350}$ hereafter) are equivalent to deeper trees (maximum number of nodes $= 500$, model RF$_{500}$ hereafter), provided that forest storage size is proportional to the nodes number.

    \subsection{CNN\label{sec:sec3-3}}
	
	We used a U-Net architecture \citep{ronneberger2015}, which is a fully convolutional network that generates images from images, the name of which comes from its U-shaped architecture in which convolutional layers are separated first with pooling layers and then with transposed convolutional layers. The first phase, with pooling layers, reduces the size of images, which is known to capture context of input images. The second phase, with transposed convolutional layers, increases the size of the contracted images, enabling precise localization. These particularities fit the needs of forecast correction.

	The architecture used is adapted from that described in \citet{ronneberger2015}. We used a padding of 1 in order to have the same resolution for inputs and outputs of the U-Net. Adding a padding generates inconsistencies on the boundaries of the patches. The input patches are then overlapped and the outputs are cropped to remove the boundaries of the output patches, resulting in $48 \times 48$ output patches from the $64 \times 64$ input patches. There is no activation function after the final $1 \times 1$ convolutional layer in order to produce a regression. Finally, the loss function used is the mean square error (MSE) since the U-Net is designed to perform a regression of the TCC value. Additional modifications are described in the next sections. We used the PyTorch library of Python for the deep learning step, and optimizations concerning the learning phase are described in \citet{kivachuk2020}.

    \subsubsection{Architecture modifications\label{sec:sec3-3-1}}
    
    \paragraph{Modified U-Net}~\\ 
    Deep learning algorithms are known to be black boxes. When many predictors are used, a first step to facilitate the interpretation of the model consists of estimating a ranking of predictors importance. Few methods exist. The sequential forward (or backward) selection is a well known method for performing such ranking. It requires however several trainings making its application to deep learning algorithms difficult. The permutation importance method, initially developed for random forest algorithms \citep{breiman2001}, was recently used to interpret CNN results in atmospheric studies \citep{mcgovern2019, toms2019}. The ranking is then performed after the training phase, and can require a large computational time.

    Selecting the most useful predictors produces a similar problem. \citet{chapados2001}, \citet{simila2007}, \citet{simila2009} and \citet{tikka2008} performed such selection for simple multilayer perceptrons. The loss function was completed by a block-penalization calculated on the weights of the first layer associated to each variable, yielding zero weights for unuseful variables. Selection was not the purpose of our work, but we developed a new method of ranking, based on a similar idea, by modifying the traditional U-Net architecture in order to let it perform its own predictors ranking during its training. 
    Before going through the U-Net, all the predictors $X$ are multiplied by a trainable weight $w$. The input of the U-Net are then $wX$. The values of the $w$ can be interpreted as coefficients of importance of each predictor (cf. section \ref{sec:sec4-4}). We preferred to add this new layer in order to have a unique weight by variable, which is easier to interpret than the hundreds of weights of the first convolutional layer would have been. The additional computational time is negligible considering that there are only $N$ additional weights to train (with $N$ the number of predictors used) which in our case is negligible in comparison to the number of weights of the U-Net itself, and that there is no modification of the loss function.

    \paragraph{Other minor modifications}~\\ 
    Some minor modifications of the U-Net architecture were tested, mostly in order to add complexity. The max-pooling layers were replaced by convolutional layers ($2 \times 2$ kernel, stride $=2$, padding $=0$, model U-Net$_{\times, \text{ down conv}}$ hereafter) which can improve results \citep{springenberg2014}. The kernel size was also increased from $3 \times 3$ to $5 \times 5$ only for the first convolutional layers before the first max-pooling layer (model U-Net$_{\times, \text{ k5}}$ hereafter).

    \subsubsection{Transformation of ground truth\label{sec:sec3-3-2}}
    
	\paragraph{Logistic transformation}~\\ 
    TCC is a bounded variable with values ranging from 0 to 100 \% with a maximum of occurrence for bound values (U-shaped distribution, cf. figure \ref{fig5}i). Two problems arise when applying machine learning methods to reproduce such variables: producing the frequent bound values and not producing values outside of the range. \citet{bottai2010} proposed the logistic transformation to deal with such variables:
	
\begin{equation}
	h\left( \text{TCC} \right) = \log \left( \frac{ \text{TCC} - \text{TCC}_{min} + \epsilon}{\text{TCC}_{max} - \text{TCC} + \epsilon} \right)
	\label{eq:logistic}
\end{equation}
	where $\text{TCC}_{min} = 0$ \%, $\text{TCC}_{max} = 100$ \% and $\epsilon = 0.001$ (higher values were also tested, model U-Net$_{\times, \text{ }\log_{0.1}}$ hereafter for $\epsilon = 0.1$) a small value determining the shape of the transformation, smaller values of $\epsilon$ producing sharper transformations. This transformation is applied to the analysis as well as the ARPEGE TCC used as a predictor for all machine learning methods (LQR, RF and CNN).
	
	\paragraph{Smoothed ground truth}~\\ 
	The analysis on which we train the U-Net contains a lot of small scale spatial variations. It is not realistic to expect the U-Net to reproduce that heterogeneity (cf. section 4.\ref{sec:sec4-3}). A solution could have been to use a generative adversarial network \citep{goodfellow2014} but it is out of the scope of the study. Moreover, these local variations can be seen as noise and disturb the learning of the model.
	
	Instead, we chose to separate this small scale heterogeneity from the large scale cloud structures. The large scale TCC is calculated by smoothing the analysis taking the median value over a square region of $0.9^{\text{o}}\times0.9^{\text{o}}$. The difference between the smoothed and the raw analysis is considered to be the small scale heterogeneity. A model was trained taking the smoothed analysis as target (model U-Net$_{\times, \text{ smooth}}$ hereafter). This allows the U-Net to focus on the representation of large scale cloud structures while not trying to reproduce the small scale variations.

    \subsubsection{Loss functions\label{sec:sec3-3-3}}
    
    As explained before, the TCC has a U-shaped distribution. The values different from 0 and 100 \% are then underrepresented, which can prevent the good representation of these values by the U-Net. A common way to balance the dataset is over (or under) sampling. It consists of duplicate (or remove) underrepresented values (overrepresented values). This is very delicate to apply to our dataset because the targets are 2D continuous data. Another common way, more adapted to our data, is the weighted loss function. It consists of increasing the importance of the underrepresented values by increasing the importance of the errors made on these values. This is done by multiplying the loss function by a weight depending on the value of the corresponding target value, the weight increasing with the rarefaction of the target value.

    Moreover, in order to improve particular aspects of the prediction, some metrics (hit rate, false alarm rate) were added to the MSE. These metrics, among other metrics used to evaluate the forecasts, are described in the next section.


    \subsection{Cloud cover forecast evaluation\label{sec:sec3-4}}
    
    The evaluation of cloud forecasts follows the World Meteorological Organization's (WMO) guidelines \citep{WMOnuages}. Firstly, they recommend that the truth and model distributions are analyzed and compared. They also recommend that data and results be stratified (lead time, diurnal cycle, season, geographical region, cloud cover threshold). We chose a threshold of 10 \% to distinguish clear to overcast skies because Earth observation from space needs very low cloud amount. Moreover, performances are calculated for 3-month seasons corresponding to meteorological seasons (Dec--Jan--Feb, Mar--Apr--May, Jun--Jul-Aug and Sep--Oct--Nov) as well as monthly. Performances are also calculated and represented as maps in addition to regional metrics (see figure \ref{fig1}) in order to perform a spatial evaluation.

    We used traditional metrics to assess continuous variables: the mean error (ME) and the mean absolute error (MAE).
    Moreover, the threshold defined above allows to evaluate the representation of the clear sky forecast (named "event" in the paragraph), based on the contingency table: the proportion correct (PC) which evaluates the good classification rate; the hit rate (HR) which is the good classification rate when the event $\text{TCC} \leq 10 \%$ was observed; the false alarm rate (F) which is the proportion of misclassification when the event was not observed; the Pierce skill score (PSS $ = $ H $ - $ F) which evaluates the overall event forecast by balancing the true-positives and the false-positives fractions; and the false alarm ratio (FAR) which represents the fraction of misclassification when the event was forecast. See \citet{wilks2011} for a detailed description of these metrics.

    We used skill scores to measure the relative improvement yielded by the CNN compared to ARPEGE. The use of skill scores is motivated by a desire to equalize the effects of intrinsically more or less difficult forecasting situations (very low cloud amount over North Africa and high amounts over the north part of the domain), when comparing forecasters or forecast systems.

    In order to evaluate the significance of the results, we performed a k-fold cross validation using four 6-month subsets (January to June and July to December for 2017 and 2018) as test data. Then, we bootstrapped each test subset to evaluate the dispersion of metrics \citep{wilks2011}. In practice, the bootstrap consisted of 30 random draws with replacement of 120 dates on each test subset, resulting in a total of 120 subsets of 120 dates each. Metrics are calculated for each subset, yielding a distribution for each metric.


\section{Results and discussion\label{sec:sec4}}
    
    \subsection{Comparison of methods\label{sec:sec4-1}}
    
	\subsubsection{Statistical comparison}
	
\begin{figure}[h]
 \centerline{\includegraphics[width=\linewidth]{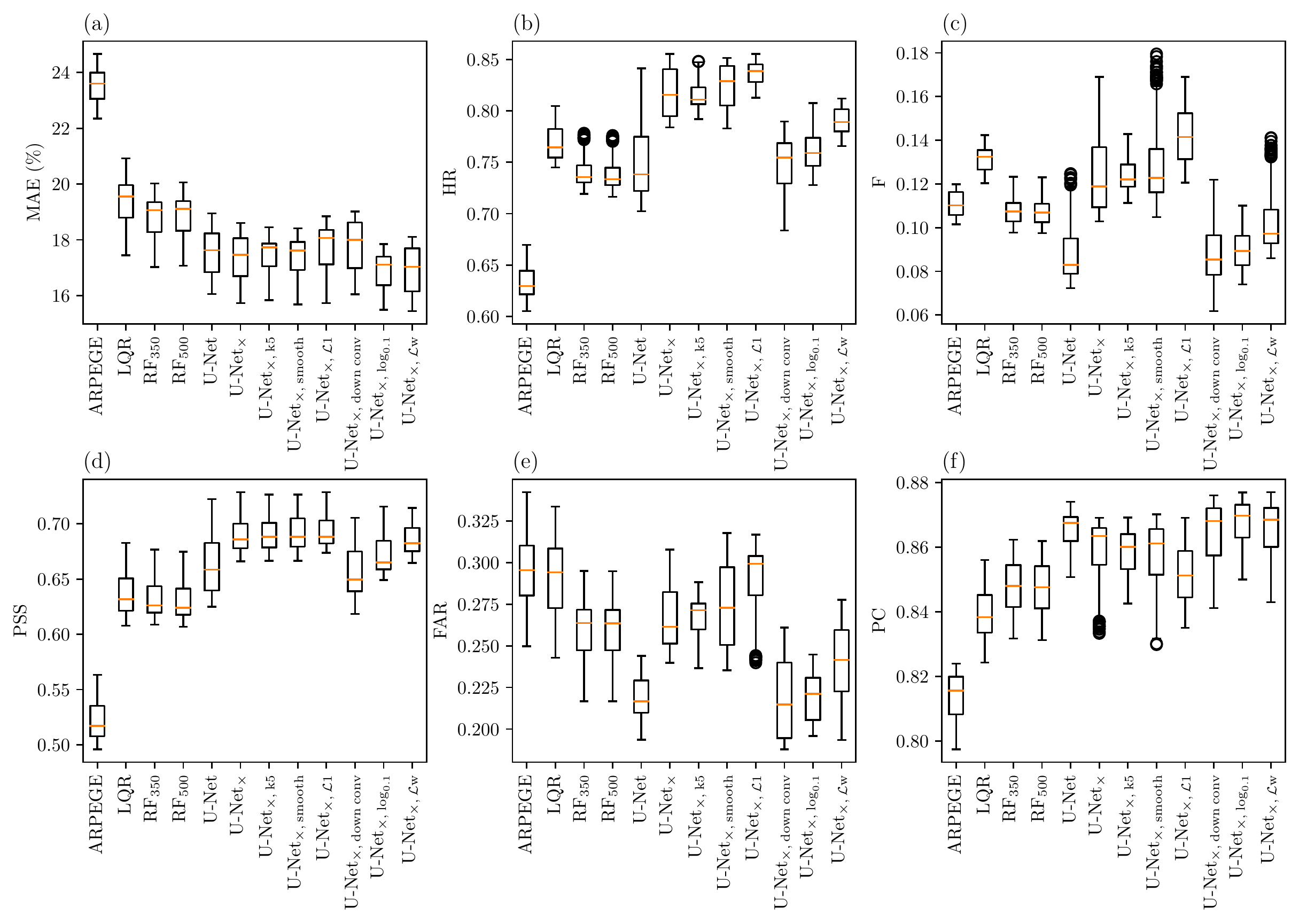}}
  \caption{Summary of performances for the ARPEGE forecast and its postprocess using LQR, RF and U-Nets, represented with boxplots. The U-Net corresponds to the traditional U-Net architecture while the "$\times$" in U-Net$_{\times}$ means that the weighted predictors layer was added. $\log_{0.1}$ corresponds to the modification of the  $\mathcal{L}_1 = \text{MSE} + 0.1 \times \left( 1 - \text{HR} \right) + \text{F}$. $\mathcal{L}_w$ corresponds to the weighted loss function where squared errors are multiplied by 3 for true TCC between 10 and 90 \%. See section \ref{sec:sec3} for a description of the other notations.}\label{fig2}
\end{figure}

	A summary of perfomances for all the postprocessing methods is given in figure \ref{fig2}. All of the different models improve most of the metrics compared to the ARPEGE forecasts, the only exception being the F score for which values increase for some models. RF is slightly better than LQR on most of the metrics, the only exceptions being the PSS for which there are not significant differences and the HR for which the LQR is better. RF's depth does not impact the performances since there are no significant differences between the RF$_{350}$ and RF$_{500}$. However, the U-Nets globally have significant better results than the LQR and RF. 
	
	The traditional U-Net architecture (UNet on the figure \ref{fig2}) is one of the models that improve the F score. However, although the HR increases compared to the ARPEGE forecasts, it is much lower than with the other models, resulting in low PSS value compared to the other U-Nets.
	The modified U-Net architecture (U-Net$_\times$ on the figure \ref{fig2}) improves most of the results. Although the F slightly increases, the PSS is much higher due to a larger increase of HR. We chose to take this architecture as baseline, before performing additional architecture modifications, because of its better global results compared to the simple U-Net, and because the multiplicative layer is needed for the ranking of variables.
	
	The other U-Net architectures can be categorized in three categories: no impact, increase or decrease of hit rate and false alarm. Using of a larger kernel for some convolutional layers (U-Net$_{\times, \text{ k5}}$) or train on a smoothed ground truth (U-Net$_{\times, \text{ smooth}}$) neither increases nor decreases metrics in a significant way. 
	HR, F and FAR decrease in U-Net$_{\times, \text{ }\log_{0.1}}$, U-Net$_{\times, \text{ }\mathcal{L}_w}$ and U-Net$_{\times, \text{ down conv}}$. As expected, the weighted loss function, with an increase of penalization for intermediate TCC values (between 10 \% and 90 \%), diminishes the absolute errors made on these values (MAE for these values drops from 35.6 \% to 32.8 \%), but increases them for other values (MAE increases from 11.5 \% to 12.3 \% for TCC $\leq 10 \%$ and from 8.5 \% to 9.0 \% for TCC $\geq 90 \%$). The TCC fields are smoother (not shown), producing a flattening/smoothing of the distribution explaining a decrease of classification metrics (except for PC). The modification of the logistic transformation also improves the representation of intermediate values by increasing the range of transformed values dedicated to these intermediate values. This has the same effect as the weighted loss function (flattening of the distribution and better representation of intermediate values balanced by increase of errors on other values). The U-Net$_{\times, \text{ down conv}}$ has the same effect for unknown reasons.
	Using a loss function combining MSE, HR and F (U-Net$_{\times, \text{ }\mathcal{L}_1}$ with the loss function $\mathcal{L}_1 =$ MSE $ + 0.1 \times \left( 1 - \text{HR} \right) + \text{F}$) causes an increase of classification metrics (except the PC and in lower proportion the PSS).
	
	There is no one U-Net that outperforms the others. The modifications, relatively to the U-Net$_\times$, either improved the regression or the detection of clear sky or the detection of high TCC values, but not at the same time. Finally, the U-Net$_\times$ has the best overall performances, balancing between detection of clear and covered sky and regression precision. We then analyse its results on the following.

    \subsubsection{Operational considerations}
    
    Implementation difficulty is crucial in operational calculations. There are two key parameters to consider: the size of the model, that has to be as light as possible, and the running time needed to process one forecast, that has to be as small as possible to ensure a quick forecast. The RFs are much heavier (some Go) than the LQR and the U-Nets (some Mo). Concerning the time of calculation, it takes only a few seconds to process an example across the whole domain using the U-Net on a gpu, which is correct considering that forecasts are for several hours ahead.

	\subsubsection{TCC fields characterisctics}
	
\begin{figure}[h]
 \centerline{\includegraphics[width=\linewidth]{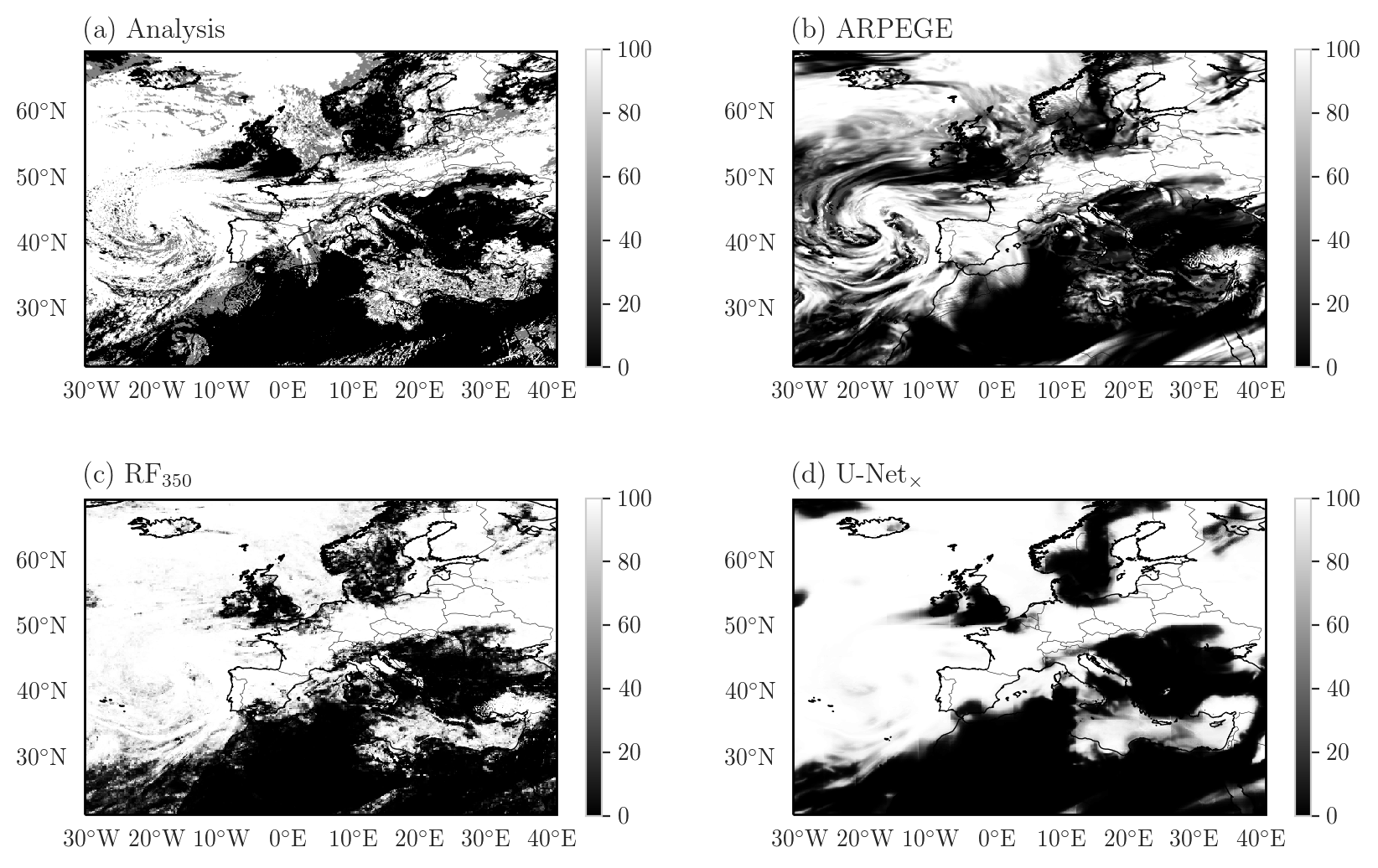}}
  \caption{Comparison of TCC values (\%) for the the 02/01/2017 for the analysis (a), ARPEGE (b), the RF$_{350}$ (c) and the U-Net$_{\times}$ (d).}\label{fig3}
\end{figure}

	The TCC fields of the analysis, ARPEGE, the RF$_{350}$ and the U-Net$_{\times}$ have all very specific particularities that make them easily recognizable (figure \ref{fig3}). Note that we only compare these two MOS methods because all the U-Nets have the same characteristics and the RFs and the LQR have the same characteristics, but RF$_{350}$ is less complex than RF$_{500}$ and reached better performances than the LQR. 
	
	In the analysis, clear sky areas have very sharp contours. Cloudy areas are either large areas of overcast, or areas of intermediate values generally characterized by an important spatial variability. 
	In ARPEGE, the TCC field is smoother, with much more intermediate values leading to an underestimation of the occurrence of overcast conditions. 
	The RF$_{350}$ has the better visual agreement thanks to a high spatial variability on some areas and a better representation of the occurrence of overcast conditions relatively to the ARPEGE forecasts. The most striking problem concerns the representation of intermediate values.
	The U-Net$_{\times}$ TCC field is very smooth, and most of the time might lead one to think of a smoothed version of the RF$_{350}$. The same problem with intermediate values occurs.
	Generally, differences between the two postprocessing models are light, concern areas of high spatial variability and are at the advantage of the U-Net. However, the spatial extension of clear sky areas is better in the U-Net$_{\times}$, which is visible over Scandinavia, over Ireland and northwest from Iceland for the 02/01/2017 on the figure \ref{fig3}.
	
\begin{figure}[h]
 \centerline{\includegraphics[width=\linewidth]{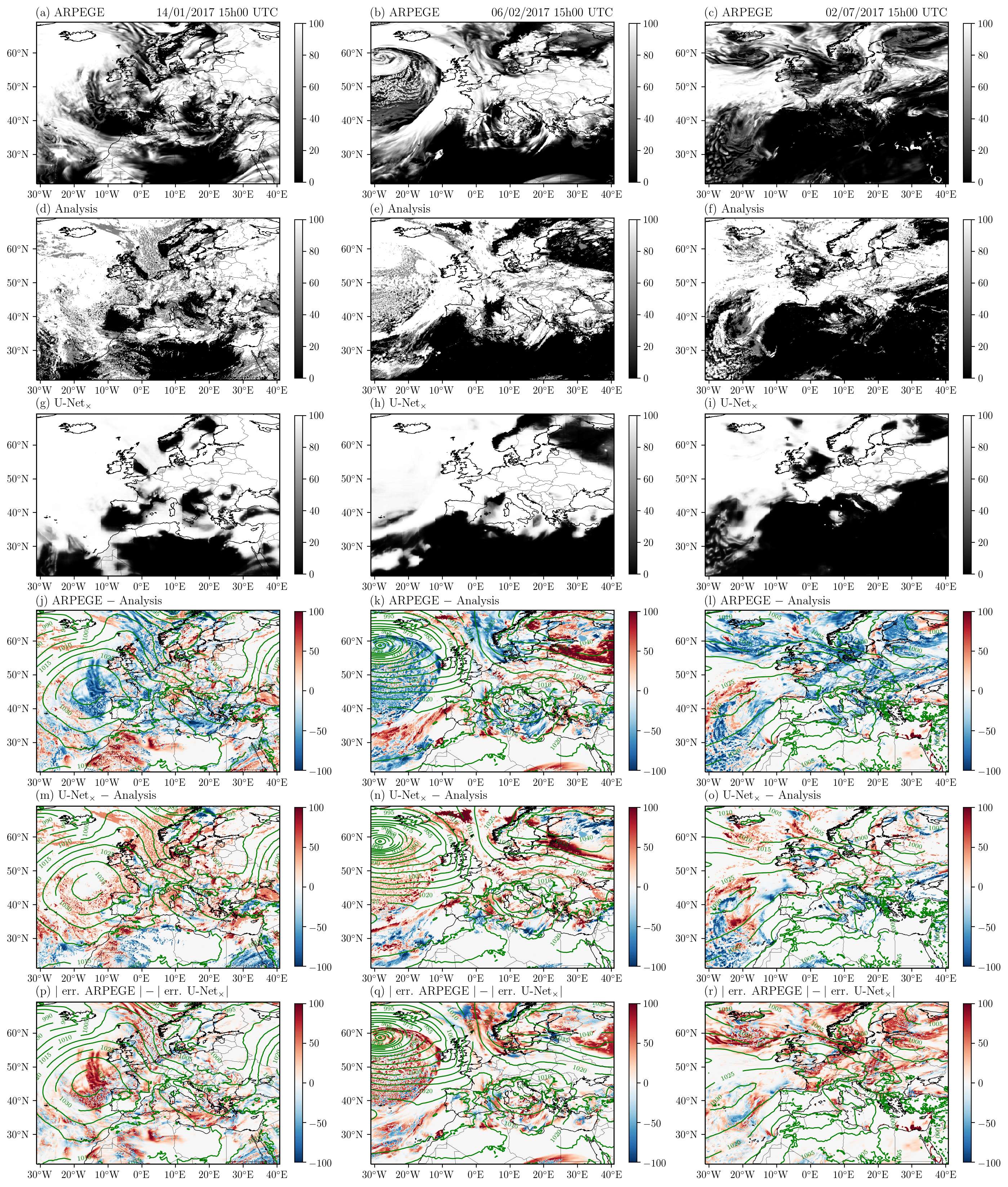}}
  \caption{Comparison between the TCC values of the ARPEGE forecasts (first row), the analysis (second row) and the U-Net$_{\times}$ outputs (third row). The forecast errors are represented on the fourth row for ARPEGE and the fifth row for the U-Net$_{\times}$, whereas the sixth row represents the improvement between ARPEGE and the U-Net$_{\times}$. On the three bottom rows, the mean sea level pressure contours are represented in green. Three situations are represented: the 14/01/2017 (left column), the 06/02/2017 (center column) and the 02/07/2017 (right column) all at 15h00 UTC. All the values are in \%.}\label{fig4}
\end{figure}

\begin{figure}[h]
 \centerline{\includegraphics[width=\linewidth]{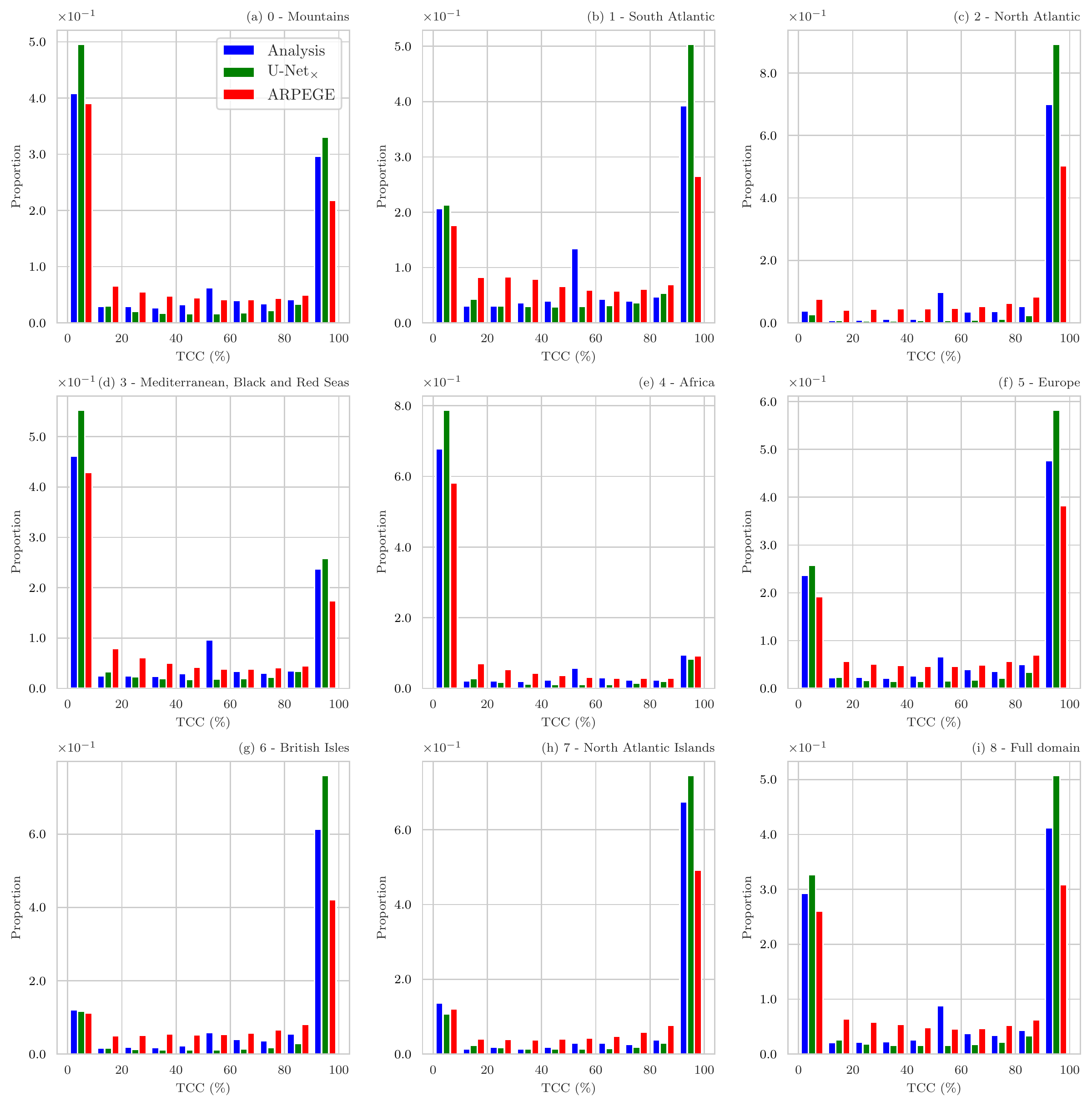}}
  \caption{Distribution of TCC per region as defined on figure \ref{fig1}, for the analysis (blue), ARPEGE (red) and the U-Net$_{\times}$ (green). The distributions on the full domain are compared in the bottom right panel.}\label{fig5}
\end{figure}

	An illustration of the improvements between the ARPEGE and U-Net$_{\times}$ forecasts is given on the figure \ref{fig4}. The situation of the 14/01/2017 perfectly highlights two key characteristics of the U-Net$_{\times}$ results: better localisation and intermediate values difficulty. The improvement on the localisation is particularly visible for clear sky areas, for which the extent was overestimated in ARPEGE, especially over the French and Spanish shore of the Mediterranean Sea where the clear sky area is very localized. Concerning cloudy areas, intermediate values are not well represented, resulting in too cloudy results for U-Net$_{\times}$, which balances with the too clear forecasts of ARPEGE. It is a recurrent bias both for the U-Net$_{\times}$ and ARPEGE for large areas of intermediate values of TCC. The situation of the 02/07/2017 is similar concerning the improvement of the localisation, leading to a better forecast of a large overcast area over Europe which was too clear in ARPEGE.

	On the 06/02/2017 at 15h00 UTC (center column of the figure \ref{fig4}), there was a low pressure system centered on the Atlantic Ocean, south from Iceland. The important cloud cover associated to that system is underestimated in ARPEGE. Too clear sky over Lows is a recurrent error in ARPEGE. The U-Net$_{\times}$ slightly overestimates the TCC on that situation, as a result of the difficulty to represent intermediate values. However, the U-Net$_{\times}$ is closer to the analysis than is ARPEGE, representing an improvement of the forecast of this situation in particular, repeated for most low pressure systems on the Atlantic (see also the figure \ref{fig3} for an other example with a low pressure system centered off Portugal). These three situations also highlight the recurrent too clear sky associated with marine clouds in ARPEGE, and the effectiveness of the U-Net$_{\times}$ to improve their forecasts.


    \subsection{Climatological and seasonal results\label{sec:sec4-2}}
    
    On the full domain, the traditional U-shaped distribution of the TCC is well marked in the analysis as well as in the ARPEGE and U-Net$_{\times}$ forecasts (figure \ref{fig5}, bottom right). In ARPEGE, there is a flattening of the distributions, for all sub-regions, resulting in an under-prediction of overcast and clear sky conditions and an over-prediction of intermediate cloud covers. \citet{crocker2013} also noticed a flattening of the distribution in the MetUM model. Overall, the U-Net$_{\times}$ corrects the forecast of occurrence of clear sky and overcast. However, the sub-region distributions reveal a tendency to over-estimate the condition with the higher occurrence: too many forecasts of clear sky over Africa and seas or too many forecasts of overcast over British Isles and the northern part of the Atlantic Ocean. It is the sign that the U-Net$_{\times}$ over-reacts to the climatic differences.
    
\begin{figure}[h]
 \centerline{\includegraphics[width=\linewidth]{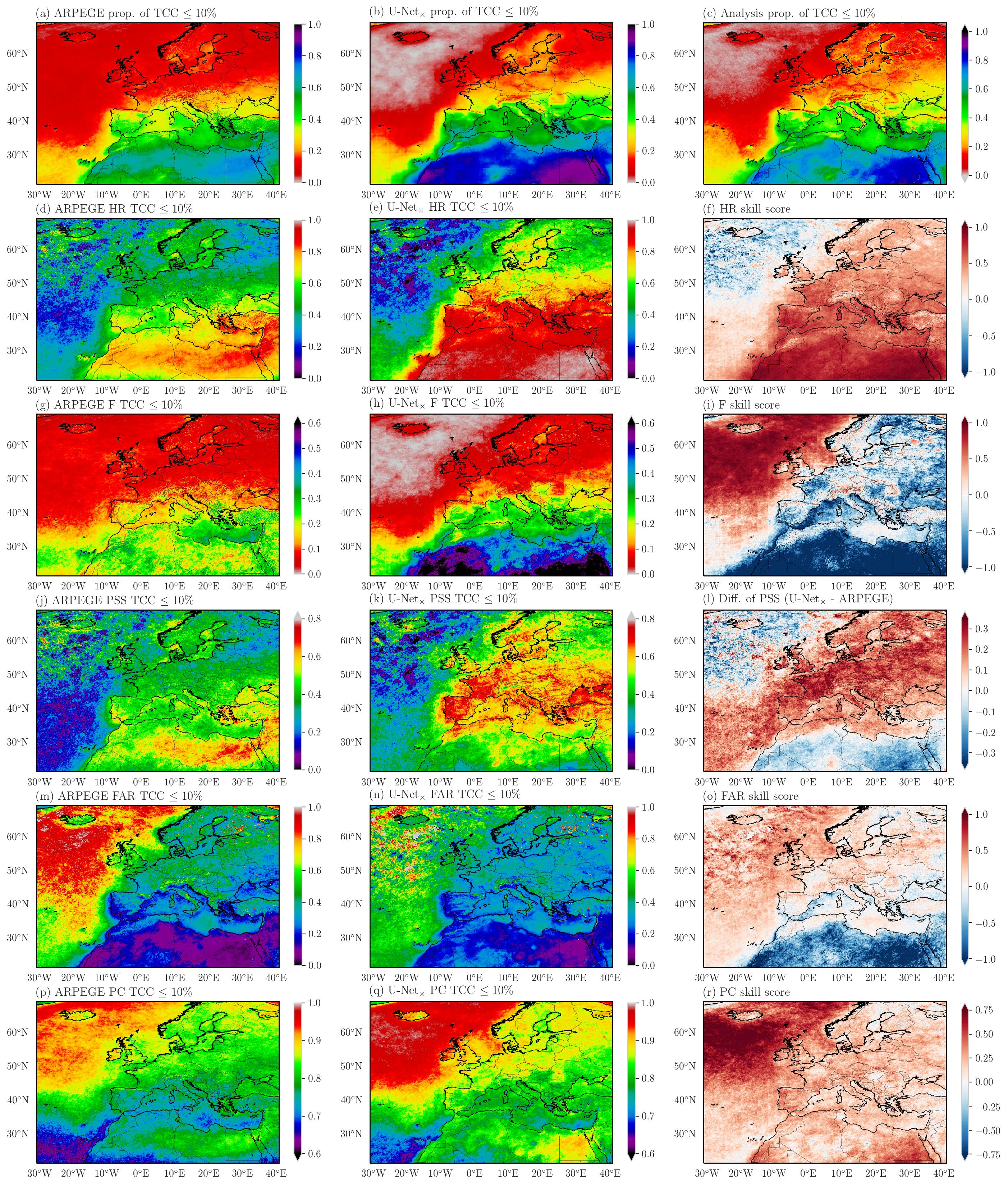}}
  \caption{Comparison of classification metrics on the 2017--2018 period between ARPEGE and the U-Net$_\times$.}\label{fig6}
\end{figure}

	The proportion of clear sky also highlights the over-representation of climatic characteristics by the CNN (figure \ref{fig6}), although the proportions are closer to the analysis than the ARPEGE forecasts. This results in an improvement of the classification (PC) skill over the entire domain (figure \ref{fig6}r), with maximum skill improvements over the northern part of the Atlantic and Egypt, corresponding respectively to the least clear and most clear regions. On the other hand, the FAR skill decreases over Africa as a result of the over-estimation of clear sky occurrence (figure \ref{fig6}o). Likewise, the F skill decreases over Africa (figure \ref{fig6}i). Over the Atlantic, the over-estimation of overcast occurrence results in a decrease of the HR skill since very few clear skies are forecasted (figure \ref{fig6}f). Overall, the prediction improves over most of the domain, except over Africa and the northern part of the Atlantic ocean (figure \ref{fig6}l).
	
\begin{figure}[h]
 \centerline{\includegraphics[width=\linewidth]{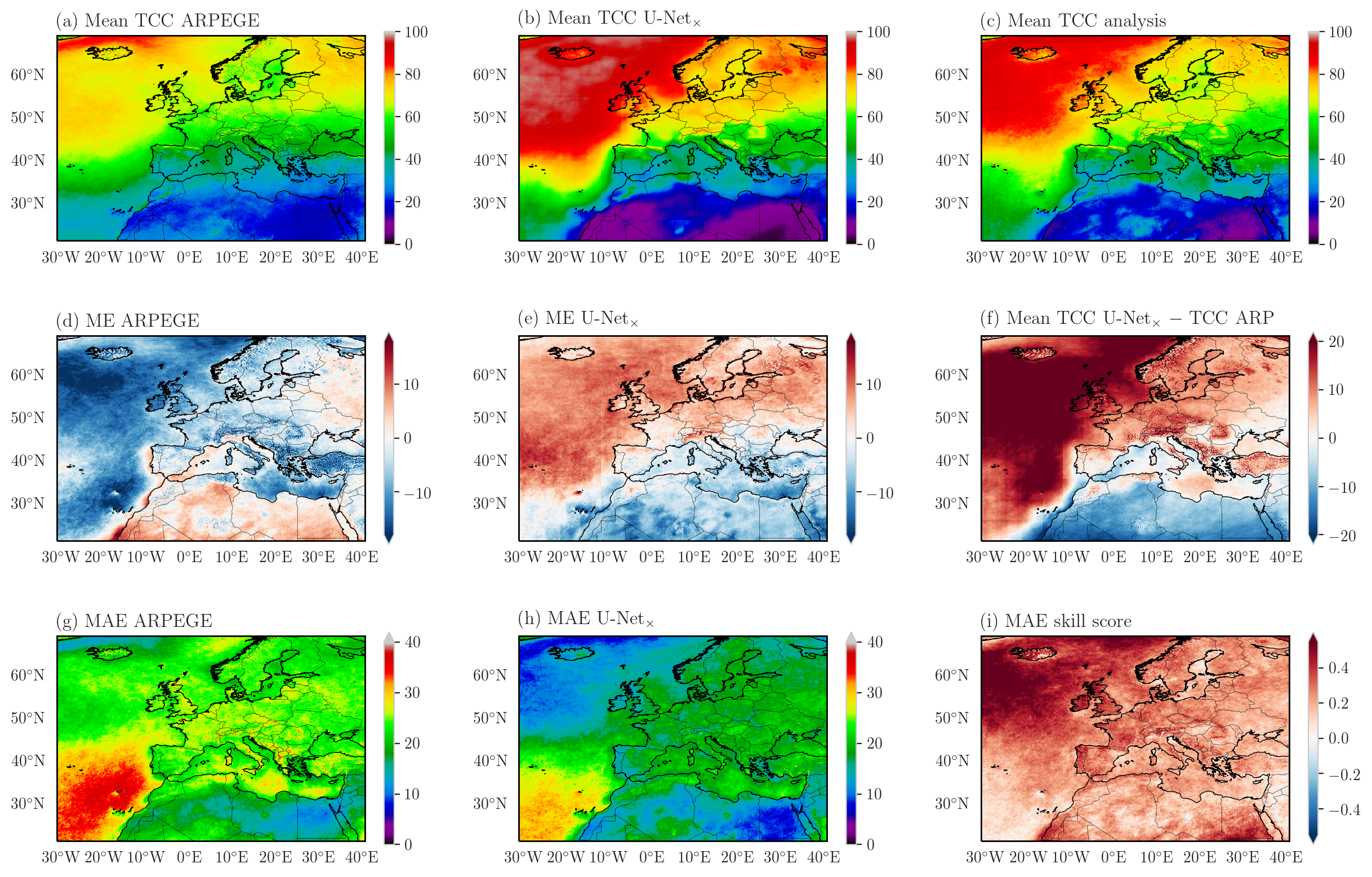}}
  \caption{Comparison of TCC ARPEGE and U-Net$_{\times}$ forecasts. The mean TCC over the 2-years period for ARPEGE (a), the U-Net$_{\times}$ (b) and the analysis (c) are compared in the top row. The middle row compares the mean errors (against the analysis) for ARPEGE (d) and the U-Net$_{\times}$ (e) while the figure (f) represents the mean difference between the U-Net$_{\times}$ and ARPEGE. The mean absolute errors of ARPEGE (g) and the U-Net$_{\times}$ (h) as well as the related skill score (i) are represented in the bottom row. All the values are in \%.}\label{fig7}
\end{figure}

	The mean TCC is also a good way with which to evaluate the climatology of the forecasts. Firstly, the latitudinal gradient, characteristic of climate differences with an increase of the values with the increase of the latitude, is well reproduced in ARPEGE (figure \ref{fig7}a). However, the maxima are not well reproduced, resulting in a positive mean deviation over Africa and mostly negative over the rest of the domain (figure \ref{fig7}d). The U-Net$_{\times}$ also reproduces the latitudinal gradient. However, as already seen before, and contrary to ARPEGE (figure \ref{fig7}b), the maxima are slightly over-estimated. There is however a better agreement with the analysis for the U-Net$_{\times}$ than for ARPEGE forecasts, which is also confirmed by the lower mean error values (figure \ref{fig7}e). The area off Africa appears to be the region with the highest errors, which was not the case with the classification metrics. This is discussed in section 4.\ref{sec:sec4-3}, as are some other strengths and limitations.
	
\begin{figure}[h]
 \centerline{\includegraphics[width=\linewidth]{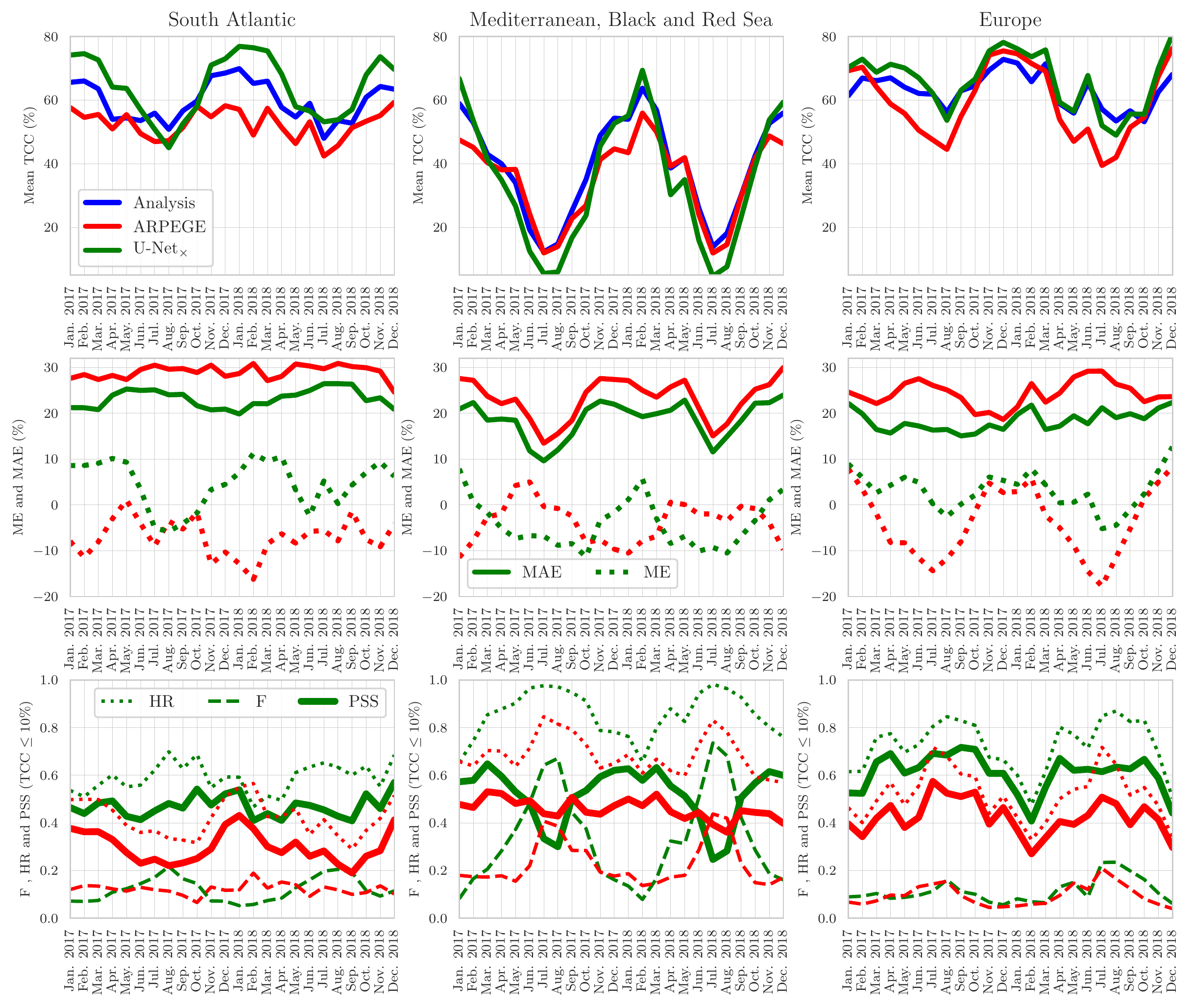}}
  \caption{Monthly metrics calculated on the South Atlantic ocean, seas and Europe as described on the figure \ref{fig1}. Colors for the figures of the second and third rows are the same as the first line, red for ARPEGE and green for the U-Net$_{\times}$}\label{fig8}
\end{figure}

	Besides regional climatological differences, the cloud cover is also marked by seasonal variations which influence forecast performances. Over the southern part of the Atlantic Ocean, over the seas of southern Europe and over Europe (we selected these regions over the eight described on the figure \ref{fig1} because they have a very clear seasonal cycle which is easier to interpret), there is a clear seasonal cycle with a maximum of cloud cover during the winter (figure \ref{fig8}). As for the representation of the climatology, the U-Net$_{\times}$ exaggerates the seasonal cycle, with an over-estimation of cloud covers during the winter and an under-estimation during the summer. It is however better than the ARPEGE forecasts, especially over the southern part of the Atlantic where the seasonal cycle is barely represented. 
	
	Classification metrics follow the same seasonal cycle, with an increase in the HR and F metrics as a result of the decrease of the mean TCC. Note that the U-Net$_{\times}$ generally improves the HR metric relative to ARPEGE (only one exception in February 2018 over the southern part of the Atlantic), while the F worsen most of the time. This is a result of the under-estimation of clear sky conditions in ARPEGE (flattened distribution) while they are over-estimated by the U-Net$_{\times}$. Indeed, the over-estimation facilitates the detection of clear sky conditions (increase of HR) but it also increases the false alarms. The PSS cycle has different specificities as it evaluates the capacity of the model to distinguish between the two classes. Its worst performances generally occur during the season with the biggest differences between the two TCC classes: minimum of cloudy conditions occurrence during the summer over the southern Europe seas and minimum occurrence of clear sky conditions during the winter over Europe. On that point, the situation of the southern part of the Atlantic is different, which can result from the higher spatial variability (cf. section \ref{sec:sec4-3}).
	
\begin{figure}[h]
 \centerline{\includegraphics[width=\linewidth]{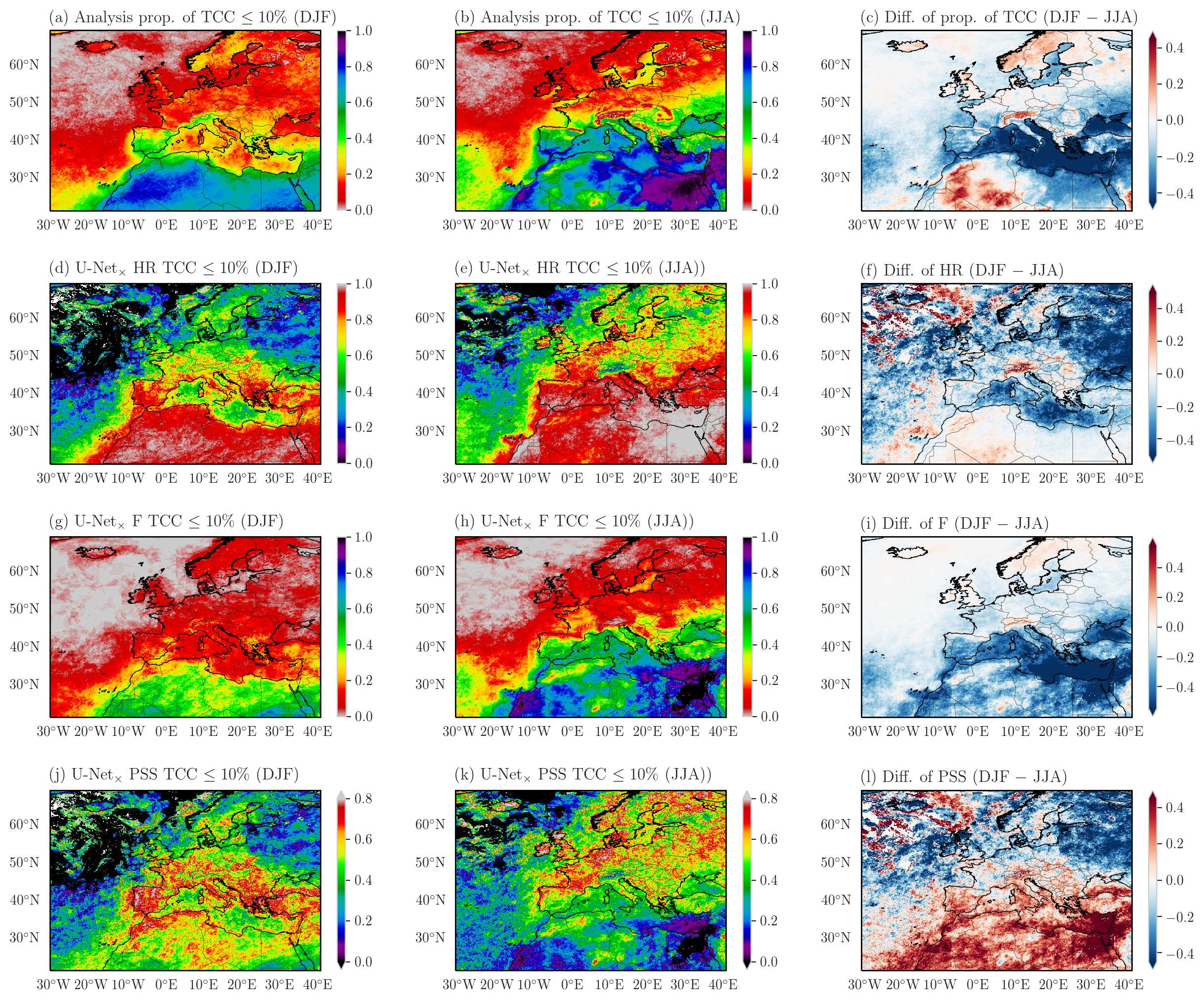}}
  \caption{Comparison of U-Net$_{\times}$ classification metrics between the winter (December to February) and the summer (June to August).}\label{fig9}
\end{figure}

	This relationship is generalized over the whole domain (figure \ref{fig9}), except that the proportion of clear sky conditions decreases during the summer over some regions such as Africa or mountainous regions.

    \subsection{Strengths and weaknesses\label{sec:sec4-3}}
	
    \subsubsection{Performances on the regression}
    
\begin{figure}[h]
 \centerline{\includegraphics[width=\linewidth]{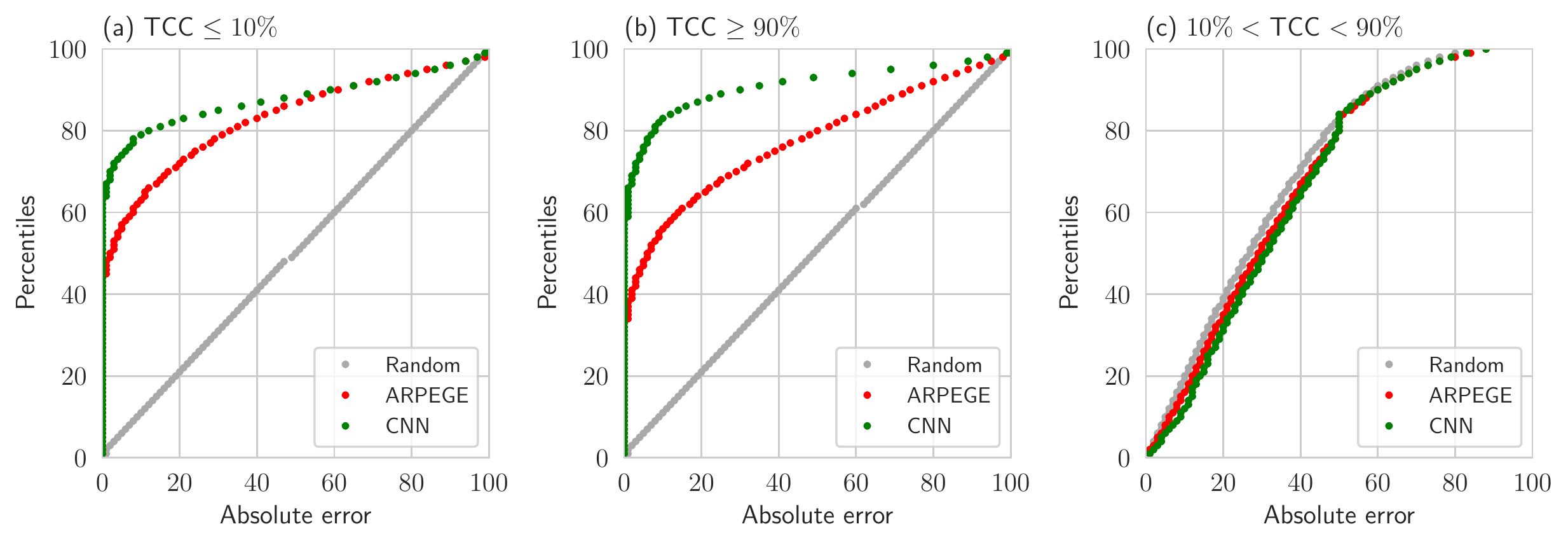}}
  \caption{Comparison of the distribution of absolute errors (in \%) in ARPEGE and the U-Net$_{\times}$ relatively to the TCC value: for $\text{TCC}\leq 10 \%$ (a), $\text{TCC}\geq 90 \%$ (b) and $10 \% < \text{TCC} < 10 \%$ (c). The errors calculated on a randomly generated dataset are in grey.}\label{fig10}
\end{figure}

    Concerning the low ($\leq 10 \%$) and high ($\geq 90 \%$) values of TCC, the U-Net$_{\times}$ improves the precision of the forecast (figure \ref{fig10}). For the low TCC values, the number of errors of magnitude lower than 50 \% decreases while it is stable for greater magnitudes. For the high TCC values, there is an improvement in the accuracy independently of the magnitude of the errors. On the contrary, there is no improvement in the accuracy concerning intermediate values of TCC. Worse, both the ARPEGE and U-Net$_{\times}$ predictions seem to have no more skill than a random forecast (in grey on the figure \ref{fig10}c). Note that the distribution of the random forecast errors does not follow the $x=y$ line because of the unbalanced TCC distribution, which produces more errors in the range 0-50 \% than in the range 50-90 \%. Intermediate values of TCC are generally related to high spatial heterogeneity, which is difficult for the U-Net$_{\times}$ to reproduce, as detailed hereafter.

    \subsubsection{Local variability}
    
    The southwestern corner of the domain, the Atlantic Ocean off Africa, is particular since regarding the mean absolute error values (figures \ref{fig7}g and h) it seems to be a challenging area, both for ARPEGE and the U-Net$_{\times}$. The low value of mean cloud cover is well reproduced however by the forecasts (figures \ref{fig7}a, b and c). 
    
    Firstly, this area approximately corresponds to the position of the North Atlantic Gyre (a clockwise-rotating system of currents in the North Atlantic), which is consistent with the results of \citet{king2013} who showed that oceanic gyres are always associated with a local minimum of cloudiness. We didn't use oceanographic data to train the CNNs. Adding oceanic current data, sea surface height or sea surface temperature (SST) could potentially help to improve the prediction over that region -- correlations have already been identified between, on one hand SST and low troposphere stratification, and on the other hand low level clouds and marine stratus and stratocumulus clouds \citep{norris1994, eastman2011}.

\begin{figure}[h]
 \centerline{\includegraphics[width=\linewidth]{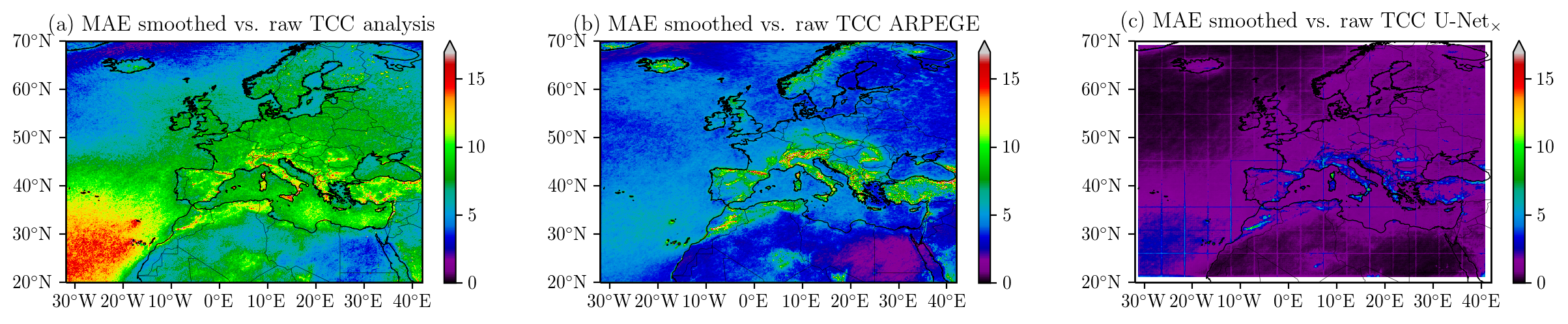}}
  \caption{Mean absolute error (in \%) calculated between the raw TCC and its smoothed version for the analysis (a), ARPEGE (b) and the U-Net$_{\times}$ (c). For each grid cell, the smoothed value corresponds to the median value over a $0.9^{\text{o}} \times 0.9^{\text{o}}$ area centered on that grid cell.}\label{fig11}
\end{figure}

    Secondly, the analysis of TCC contains some local high spatial variability areas (mackerel sky, marine stratocumulus clouds for example). We define the variability as the difference between the "raw" and a smoothed version of the analysis, as described in the section 3.c.\ref{sec:sec3-3-2}. A climatology of these variabilities is represented in figure \ref{fig11}. Although the North Atlantic Gyre area is the most heterogeneous area, both ARPEGE and the U-Net$_{\times}$ are unable to reproduce that, explaining the lower precision of calculations. The figures \ref{fig3} and \ref{fig4} illustrate that lack of spatial variability in the U-Net$_{\times}$ and in a less extent in ARPEGE. Moreover, the comparison of the variability with the MAE of the ARPEGE forecasts (figure \ref{fig7}g) shows a high correlation with an increase of MAE with the increase of variability. This is even more obvious for the MAE of the U-Net$_{\times}$ predictions (figure \ref{fig7}h). 

    Thirdly, the proportion of intermediate values of TCC (between 10 and 90 \%) is higher in that region (figure \ref{fig5}). However, as detailed before, the U-Net$_{\times}$ obtained its worst results on these values (figure \ref{fig10}). Concerning marine stratocumulus clouds (MSC), they are very sensitive to the aerosols load in the atmosphere, a high amount leading to closed cells for which the TCC is generally close to 100 \%, whereas lower amounts lead to open cells that have typical TCC less than 65 \% \citep{wood2008, wood2011}. Adding aerosol content data could therefore help differentiate these two regimes of MSC for a better representation of intermediate values and the associated variability.

    \subsubsection{Mountains}
    
    Mountainous regions (the Alps, the Asturies, the Atlas, the Balkans, the Carpathian Mountains, the Italian peninsula, the Massif Central, the Pyrenees, \ldots) present interesting local patterns with an increase of the mean TCC in comparison with the values of the surrounding regions (figure \ref{fig7}c), also visible with a decrease of the clear sky occurrence (figure \ref{fig6}c). This is in agreement with \citet{barry2008} which details that cloud cover over mountainous regions is generally thicker and has a higher occurrence.

    Complex terrain areas are known to be challenging for weather forecast due to the misrepresentation of topography and use of inappropriate parameterizations \citep{goger2016}, especially for global models and their coarse resolution. This is confirmed in the ARPEGE forecasts with an underestimation of the mean TCC over mountainous regions, resulting in local decrease of ME and local increase of MAE. \citet{vionnet2016} also reported an underestimation of cloud cover over the French Alps using the high resolution model AROME.

    We evaluate the TCC forecasts using an analysis based on satellite observations, which can meet difficulties over highly reflective surfaces, such as snow cover over mountains during winter. However, a seasonal evaluation reveals that there is an increase of forecast errors during the summer (and in lower proportions during the spring) correlated with an increase of the underestimation of the mean cloud cover forecast (not shown). This is associated with an increase in the convective clouds amount which are clearly underpredicted in ARPEGE.

    Globally the U-Net$_{\times}$ reproduces well the local maximum of mean TCC over mountainous terrains resulting in local high skill score values. This shows that the U-Net$_{\times}$ has integrated this geographic feature and is able to handle the mountainous terrain forecasts limitations.

    
    \subsection{Predictors importance\label{sec:sec4-4}}
	
\begin{figure}[h]
 \centerline{\includegraphics[width=\linewidth]{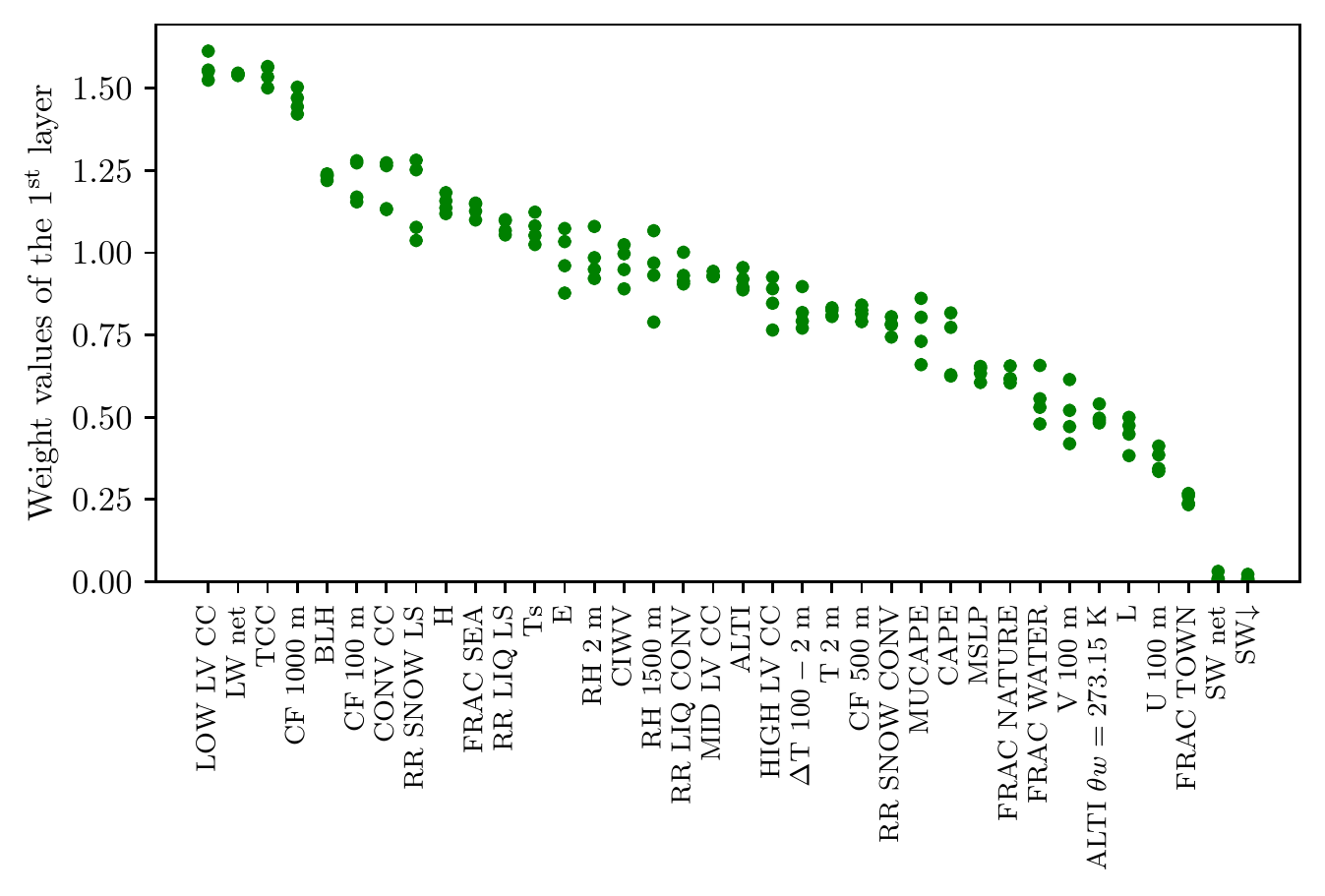}}
  \caption{Values of weight of the first layer added before the U-Net$_{\times}$. For each variable, the 4 values correspond to the 4 models trained for the cross validation. The variables are sorted by descending mean value. A description of the variables is given in the table \ref{tab1}.}\label{fig12}
\end{figure}

	The modified U-Net architecture we used, in which before going through the U-Net, each predictor is multiplied by a weight (U-Net$_{\times}$), produced better results than the traditional U-Net. The values of these weights are presented in figure \ref{fig12}. There is a clear ranking of values, giving a relative importance of each variable, which appears to help the U-Net$_{\times}$ learning. The net ordering of values makes the ranking resulting from the U-Net$_{\times}$ clear.

	\subsubsection{Cloud-related variables}
	
	Three kind of cloud-related variables were used: the TCC, cloud covers (CC) for specific conditions or atmospheric layer and cloud fractions (CF) at different altitudes. Five of the seven most important variables are directly related to clouds. It is obvious that the TCC is very important since it is the value we try to correct. The CF at 500 m contains some redundant information with CFs at 100 and 1000 m, as demonstrated by the correlation coefficients $R$: $R_{100/500} = 0.49$, $R_{500/1000} = 0.59$, $R_{100/1000} = 0.28$.
	
	Although the CC calculated for the lowest part of the atmosphere (LOW LV CC) is used to calculate the TCC, it is an important predictor. Low level clouds representation in NWP is generally challenging, making the variable possibly very inaccurate. In ARPEGE, there is a recurrent underestimation of MSC that leads to important underestimations of TCC over the Atlantic, and the same underestimation occurs over land. The CNN probably uses the low level CC to correct these errors, that it does correct most of the time, hence the importance of CC at this level despite the forecast errors. The same forecast difficulties concern the convective CC which is one of the most important predictors (CONV CC, 7\textsuperscript{th} predictor in the ranking) in contrast to CC for the middle (MID LV CC, 18\textsuperscript{th}) and the high (HIGH LV CC, 20\textsuperscript{th}) part of the atmosphere. It is not clear how convective clouds can help, but it is likely that some important forecast errors, on this variable which is also very challenging, can help the same way low level clouds do.
	
	Finally, even if it is not directly related to them, clouds affect the LW net radiation (LW net) by blocking the outgoing radiations, which can explain its importance (2\textsuperscript{nd} predictor in the ranking).

	\subsubsection{Precipitations variables}
	
	After cloud-related variables, some precipitation variables appear to be important, which makes sense given the fact that there is no rain without clouds. Large scale precipitations (RR SNOW LS and RR LIQ LS) are more important than convective ones (RR Snow CONV and RR LIQ CONV). When large scale rainfall amount exceeds at least 1 mm over 3 hours, most of the TCC of the analysis reach 100 \% (92 \% of values). This makes large scale precipitation a good predictor with which to diagnose the occurrence of very cloudy sky, which the U-Net$_{\times}$ kept since 99 \% of the TCC associated with rainfall amount exceeding that threshold reach 100 \%. For the same threshold, only 60 \% of the values (analysis) reach a TCC of 100 \% for convective precipitations. The U-Net$_{\times}$ also kept that correlation since it mainly produces overcast situations, but it is less determinant since a large part of these data concern lower TCC. 
	
	Several reasons can explain the differences between large scale and convective precipitations.	It is well known that the representation of convective clouds is a challenging task for NWP. Their extension is limited in space and in time which complicates even more their localisation with precision. On a 0.1$^{\text{o}}$ resolution grid, it is then possible that a fraction of the grid cell remains clear, the associated TCC being then lower than 100 \%. Large scale precipitations are generally associated with large cloud structures (stratiform clouds), for which the TCC values definitely reach 100 \%.
	
	Moreover, we used precipitation amounts over the previous 3 hours. Concerning convective precipitations, it is likely that precipitations were concentrated at the beginning of those 3 hours and that the sky has already started to clear. The large extent of cloud structures associated with large scale precipitations is less sensitive to that phenomenon.
	
	We attempted to see whether or not the U-Net$_{\times}$ reacts directly to the value of precipitation. During the test step, large scale precipitation values lower than 1 mm were enhanced to 1 mm. Despite non-linearities, knowing that this threshold is generally associated to overcast conditions, we expected the TCC to increase. The opposite occurred however with a diminution of TCC. The modifications on the precipitations smoothed the field, leading to the reduction of the gradients. This suggests that the CNN focuses on the spatial structures of precipitation areas more than on the values.


\section{Conclusion\label{sec:sec5}}
    
    Although CNNs are becoming the most popular deep learning tool, and despite their specialization for extracting spatial information, they are still rarely used in atmospheric sciences. We applied CNNs to postprocess the TCC forecasts of the ARPEGE model and evaluated their ability by comparison with approved machine learning techniques traditionally used in NWP postprocessing: random forests and a logistic quantile regression. The LQR and RF have attained similar performances, but the LQR is operationally easier to implement, and the CNNs are significantly better than the LQR and the RFs. 
	
	The main difference between RF's and CNN's results concerns the representation of local spatial variability. The RF reproduces this whilst the CNN produces very smoothed TCC fields, which seems to prove an advantage. Moreover, the CNN locates areas of overcast and clear sky with better precision. This demonstrates the ability of CNNs to improve NWP outputs. 
	
	The CNN has trouble predicting intermediate TCC values with precision. There is no more forecast skill than a random forecast for these values, which does not improve the ARPEGE forecasts. On the other hand, the CNN improves forecasts over mountainous regions, where errors were large in ARPEGE. The CNN also corrects the recurrent too clear sky of ARPEGE forecasts over low pressure systems, as well as those linked to low level cloud errors, although it generally produces too cloudy a sky. This kind of over-correction is also visible on the climatological and seasonal scale. The CNN exaggerates local maxima of mean TCC, leading to difficulties in detecting cloudy conditions over Africa and clear sky conditions over the northern part of the Atlantic. The amplitude of the seasonal cycle of the mean TCC was generally underestimated in ARPEGE, and although it is overestimated with the CNN, its representation is better. Performances of the CNN, such as ARPEGE performances, are then impacted by the seasonal cycle and climatological differences.
	
    We introduced a novel method for the ranking of predictors by importance. Contrary to traditional methods, such as permutation importance and sequential selection, the ranking is performed during one unique training and is negligible in additional computational time. It consists of a weighting predictor layer prior to the traditional U-Net architecture. As expected, cloud-related variables are very important. Besides the ARPEGE TCC, that is to be corrected, low level CC is very important despite their common forecast errors. Actually, it seems that these errors are useful for the CNN, especially because of the recurrent too clear sky forecasts associated with low level cloud. Besides, large scale precipitations are found to be more important than convective precipitations. It is not clear however how the CNN uses that information.

\section*{acknowledgments}
This work is part of the Deep4cast project (\footnote{https://www.researchgate.net/project/Deep4Cast}), funded by the STAE Fundation.

%
%


\bibliographystyle{ametsoc2014}
\bibliography{references}


\begin{table}[h]
    \caption{List of ARPEGE and SURFEX variables used as predictors in that study.}\label{tab1}
	\begin{center}
	\begin{tabular*}{\textwidth}{p{0.97\textwidth}}
        \textbf{Fundamental meteorological variables:} \\
        Ts: surface temperature; T 2 m: 2-m temperature; MSLP: mean sea level pressure; U and V 100 m: zonal and meridional wind components at 100 m a.g.l.  \\
        
        \textbf{Cloud-related variables:} \\
        LOW LV CC: low level cloud cover; MID LV CC: middle level cloud cover; HIGH LV CC: high level cloud cover; CONV CC: convective cloud cover; TCC: total cloud cover; CF: cloud fraction. \\
        
        \textbf{Precipitation variables:} \\
        RR corresponds to 3-hour rainfall accumulation, SNOW and LIQ distinguish snow and liquid precipitations while LS and CONV means large scale and convective precipitations. \\
        
        \textbf{Flux variables:} \\
        LW net: net longwave radiation at the surface; H: sensible heat flux; E: evaporation flux; L: latent heat flux; SW net: net shortwave radiation at the surface; SW$\downarrow$: ongoing shortwave radiation at the surface. \\
        
        \textbf{Atmospheric stability:} \\
        BLH: boundary layer height; $\Delta$T 100 – 2 m: vertical difference of temperature between 100 and 2 m; CAPE: convective available potential energy in the model; MUCAPE: most unstable CAPE. \\
        
        \textbf{Other variables:} \\
        CIWV: column integrated water vapor; ALTI $\theta w = 273.15 K$: altitude of the 0$^{\text{o}}C$ wet-bulb potential temperature level. \\
        
        \textbf{Terrain variables:} \\
        ALTI: altitude; FRAC SEA, NATURE, WATER and TOWN: grid cell fraction occupied by seas and oceans, natural surfaces, continental water bodies and artificial surfaces (from SURFEX). \\

	\end{tabular*}
	\end{center}
\end{table}

\end{document}